\documentclass[aps,pra,twocolumn,showpacs,floatfix]{revtex4}
\usepackage{graphicx}% Include figure files
\usepackage{dcolumn}% Align table columns on decimal point
\usepackage{amsmath}

\def\etal{\emph{et~al.}}
\newcommand{\eref}[1]{(\ref{#1})}
\newcommand{\Eref}[1]{Eq.~(\ref{#1})}

\newcommand{\cm}{cm$^{-1}$}
\begin{document}

\title{Calculation of the dependence of transition frequencies on the fine structure constant and the search for variation of $\alpha$ in QSO spectra}
\author{J. C. Berengut}
\author{V. A. Dzuba}
\email{V.Dzuba@unsw.edu.au}
\author{V. V. Flambaum}
\email{V.Flambaum@unsw.edu.au}
\author{M. V. Marchenko}
\affiliation{School of Physics, University of New South Wales, 
Sydney 2052,Australia}

\date{\today}

\begin{abstract}

We calculate the dependence of 
 atomic transition frequencies on the fine structure constant,
 $\alpha=e^2/\hbar c$, for some ions of Ti, Mn, Na, C, and O.
The results of these calculations will be used in the search 
for variation of $\alpha$ in quasar absorption 
spectra.

\end{abstract} 

\maketitle

The possibility that the fundamental constants vary is suggested by theories
unifying gravity with other interactions (see, e.g. \cite{theo1,theo2,theo3}
and review \cite{uzan}).
The analysis of quasar absorption spectra by means of the many-multiplet
method reveals anomalies which can be
interpreted in terms of varying fine structure constant $\alpha$
\cite{quasar1,quasar2,quasar3}.
The first indication that $\alpha$ might have been smaller at early epoch came 
from the analysis of magnesium and iron lines \cite{quasar1,quasar2}. 
Later inclusion of other lines belonging to many different atoms and ions 
(Si, Cr, Ni, Zn, etc.) as well as many samples of data from different
gas clouds not only confirmed the initial claim, but made it even stronger
\cite{quasar3}. However, there are some recent works in which a similar
analysis indicates no variation of $\alpha$ in quasar absorption
spectra \cite{Quast,Srianand}. These works use the same many-multiplet
method and the results of our calculations of the relativistic effects
in atoms, but analyze different samples of data from a different telescope. 
It is important to
include as much data as possible into the analysis to resolve the
differences, and to verify or discard the claim of a varying fine structure
constant.

It is natural to analyze fine structure intervals in the search of variation 
of $\alpha$. Indeed, initial searches of variation of $\alpha$ in quasar
absorption spectra were based on alkali-doublet lines (alkali-doublet method)
\cite{AD1,AD2,AD3} and on the fine structure of O~III \cite{Bahcall}.
However, all of the present evidence for varying fine structure constant
has come from the analysis of the $E1$-transition frequencies (many-multiplet 
method) rather than fine structure intervals. 
These frequencies are about an order of magnitude more
sensitive to the variation of $\alpha$ \cite{quasar2}. However, the corresponding
analysis is much more complicated. One needs to perform accurate 
\emph{ab~initio} calculations of the atomic structure to reveal the
dependence of transition frequencies on the fine structure constant. We have done such 
calculations for many atoms and ions in our previous works 
\cite{Dzuba1,Dzuba2}.
In the present work we do similar calculations for some other atoms and ions
for which data on quasar absorption spectra are available \cite{Murphy},
and for which corresponding calculations have not previously been done.

We use the relativistic Hartree-Fock (RHF) method as a starting point of our 
calculations. Correlations are included by means of configuration-interaction
(CI) method for many valence electron atoms, or by the many-body perturbation
theory (MBPT) and Brueckner-orbital method for single valence electron 
atoms.
The dependence of the frequencies on $\alpha$ is revealed by varying $\alpha$
in computer codes.

The results are presented in the form
\begin{equation}
        \omega=\omega_{0}+qx ,
\label{omega}
\end{equation}
where $ x = (\alpha^{2}/\alpha^{2}_{0})-1$,
$\alpha_0$ is the laboratory value of the fine structure 
constant, $\omega$ and $\omega_{0}$ are the frequencies of the transition in
quasar absorption spectra and in the laboratory, respectively, and $q$ is 
the relativistic energy shift that comes from the calculations.
Comparing the laboratory frequencies, $\omega_0$, with those measured in
the quasar absorption spectra, $\omega$, allows one to obtain the value
of $\alpha$ billions of years ago.

The method of calculations is described in detail in our early works
\cite{Dzuba1,Dzuba2}. Here we only discuss the details specific for current
calculations.

Some atoms and ions considered in the present work represent open-shell (many valence electron)
systems. Therefore, the Hartree-Fock procedure needs to be further specified.
The natural choice is to remove all open-shell electrons and start the
RHF calculations for the closed-shell core. 
However, this usually leads to poor convergence of the subsequent CI method. 
Better convergence can be achieved using the so called $V^{N-1}$ approximation
in which only one valence electron is removed.
Since we calculate not only the ground state but also excited states of different
configurations, it is convenient to remove the electron
which changes its state in the transition. Single-electron basis states
for valence electrons are calculated in the $V^{N-1}$ potential of the 
frozen-core.

The $V^{N-1}$ potential corresponds to an open-shell system.
We include the contribution of the open shells
into the RHF potential as if they were totally filled and then
multiply them by a weighting coefficient. Note that this procedure must
not destroy the cancellation of the self-action (we would like to remind the
reader that there is exact cancellation between direct and exchange 
self-action in the RHF equations for the closed-shell systems).

For the CI calculations we use a B-splined single-electron basis set similar 
to those developed by Johnson~\etal \cite{BS1,BS2,BS3}.
The main difference is that we use the open-shell RHF Hamiltonian described 
above to calculate the B-splined states.

There are two major sources of inaccuracy in the standard CI calculations. 
One is incompleteness of the basis set for valence electrons, and another is core-valence correlations.
We use a fitting procedure to model both effects.
We add an extra term into a single-electron part of the Hamiltonian for
the valence electrons:
\begin{equation}
  U(r)=-\frac{\alpha_c}{2\left(r^4+a^4 \right)}.
\label{fit}
\end{equation}
Here $\alpha_c$ is the polarizability of the atomic core and $a$ is a cut-off
parameter that is introduced to remove the singularity at $r=0$. 
We use $a=a_b$ (Bohr radius) and treat $\alpha_c$ as a fitting parameter.
The values of $\alpha_c$ for each partial wave ($s,p,d$) are chosen to fit
the experimental energy levels of the many-electron atom.

The term \eref{fit} describes polarization of the atomic core by valence 
electrons. It can be considered as a semi-empirical approximation to
the correlation interaction of a particular valence electron with the core.
It also allows us to improve the convergence of the CI calculations by
modifying the single-electron basis states. Our calculations for rare-earth
ions \cite{Safronova1,Safronova2} have demonstrated that using this term 
allows one to obtain 
good accuracy of calculations with the minimum number of single-electron
basis states (one in each partial wave in the cited works).

Below we present the details and results of calculations for the atoms and ions 
considered.
All transition frequencies are presented with respect to the ground state.
Therefore we use the term ``energy levels'' instead. If a transition
between excited states is needed, the corresponding relativistic energy
shift $q$ is the difference between the level shifts
($q_{2 \rightarrow 1} = q_2 - q_1$).

\paragraph{Manganese ($Z=25$):}  The ground state of Mn$^+$ is 
$3d^{5}4s \ ^7S_{3}$ and we need to consider transitions into the $3d^{4}4s4p$
configuration. Earlier we also considered transitions to the states of the
$3d^54p$ configuration \cite{Dzuba1}. Since in the present work we use a
different basis set, we have repeated calculations for this configuration in order
to check their accuracy.

The RHF calculations are done in the $V^{N-1}$ approximation with the
$3d^5$ configuration of external electrons. The $4s, 4p$ and higher states are
calculated in the same $V^{N-1}$ potential. We use $\alpha_c = 2.05 a_B^3$
for the $p$-wave as a fitting parameter in \Eref{fit}.
The results are presented in Table \ref{Mn}.

\begin{table}[htb]
\caption{Energies and relativistic energy shifts ($q$) for Mn$^+$.}
\label{Mn}
\begin{ruledtabular}
\begin{tabular}{llcccrlr}
\multicolumn{2}{c}{State} & \multicolumn{3}{c}{$\omega_0$ (\cm)}& 
\multicolumn{3}{c}{$q$ (\cm)} \\
&&\multicolumn{2}{c}{theory} & experiment & \\
&&no fitting & fitted & \cite{Moore} & \multicolumn{2}{c}{this work} & \cite{Dzuba1} \\ 
\hline
 $3d^5 4p$   & $^7P^o_2$ & 36356 & 38424 & 38366 &   869 & (150) &  986 \\
 $3d^5 4p$   & $^7P^o_3$ & 36519 & 38585 & 38543 &  1030 & (150) & 1148 \\
 $3d^5 4p$   & $^7P^o_4$ & 36755 & 38814 & 38807 &  1276 & (150) & 1420 \\
 $3d^4 4s4p$ & $^7P^o_2$ & 84092 & 83363 & 83256 & -3033 & (450) & \\
 $3d^4 4s4p$ & $^7P^o_3$ & 84319 & 83559 & 83376 & -2825 & (450) & \\
 $3d^4 4s4p$ & $^7P^o_4$ & 84619 & 83818 & 83529 & -2556 & (450) & \\
\end{tabular}
\end{ruledtabular}
\end{table}

We have done several tests to obtain the error estimates.
Fitting changes both energies and $q$-coefficients by less than 6\% for all 
transitions, and the agreement with previous calculations is within 15\%.
We note that the accuracy of the fine structure splitting is significantly worse than this, especially for the upper levels.
In some sense this is unsurprising: the fine structure splitting is much smaller than the total relativistic effect due to the strong cancellation between levels. Nonetheless, we may try to fit the splitting using some parameter, and see how $q$ changes. In fact we have done this by varying $\alpha$ itself. The splitting for the higher levels becomes very close to experiment at approximately $x=-0.5$, and at this point $q$ is within 1\% of the value at $x=0$. This is a similar error estimation procedure as that used in \cite{Dzuba2} for Ni~{\small II}, and it shows that $q$ is insensitive to the fine structure splitting. In another test, we have fit the fine structure splitting by introducing screening parameters before the Coulomb integrals in the CI code. This models the effect of screening of the valence electrons by the core electrons. Again, $q$ changes by less than 15\%.

More worrying is the possibility of level misidentification. This can occur due to misidentification in the tables, or alternatively in the CI results. The Mn$^+$ spectra contains two $3d^44s4p \ ^7P^o$ multiplets in close proximity. It is possible (although unlikely) that they are swapped in the computer code. The $q$ values for the other multiplet are approximately 15\% smaller than for the required transitions.
Therefore, we use 15\% as a conservative estimate of the accuracy of $q$.

Note that the relativistic
shift is positive for the $s-p$ singe-electron transitions and negative
for the $d-p$ transitions. Having transitions with different signs of $q$-coefficients
in the same atom (ion) helps to fight
systematic errors in the search for variation of $\alpha$ \cite{Dzuba1}.

\paragraph{Titanium( $Z=22$):} We perform calculations for both Ti$^+$ 
and Ti$^{2+}$ starting from the same RHF approximation, and 
using the same single-electron basis set.
The ground state of Ti$^+$ is $3d^{2}4s \ ^4F_{3/2}$
and we need to consider transitions into states of the $3d^{2}4p$
and $3d4s4p$ configurations.
The ground state of Ti$^{2+}$ is $3d^{2} \ ^3F_{2}$
and we need to consider transitions into the states of the $3d4p$
configuration.
Therefore it is convenient to do the
RHF calculations for the Ti$^{2+}$ ion
with the $3d^{2}$ open-shell configuration. The $4s$, $4p$ and other
basis states for the CI method are calculated in the frozen-core field 
of Ti$^{2+}$. 

The fitting parameters chosen are $\alpha_c=0.38 a_B^3$ for 
$s$-electrons and $\alpha_c=0.065 a_B^3$ for $d$-electrons.
The results are presented in Table \ref{Ti}.
As in the case of Mn$^+$, there are negative and positive relativistic shifts.
The effects of fitting and change of basis set does not exceed 10\%.
The values of the $q$-coefficients for titanium are consistent with 
calculations for other atoms and with semi-empirical estimations
using the formulas presented in \cite{Dzuba1}. 
In particular, the values of the negative $q$-coefficients for 
the $d-p$ transitions are very close to the values for similar transitions in 
Cr~{\small II} \cite{Dzuba1}. The positive coefficients for Ti$^+$ are very close 
to those for Mn$^+$ after rescaling by $Z^2$ according to the semi-empirical
formula \cite{Dzuba1}. The $q$-coefficients for the $3d^24p \ ^4F^o$ states closely match independent calculations of \cite{kozlov}.

\begin{table}[htb]
\caption{Energies and relativistic energy shifts ($q$) for Ti$^+$ and Ti$^{2+}$.}
\label{Ti}
\begin{ruledtabular}
\begin{tabular}{llcccc}
\multicolumn{2}{c}{State} & \multicolumn{3}{c}{$\omega_0$ (\cm)} & $q$ (\cm) \\
& & \multicolumn{2}{c}{theory} & experiment   & \\
& & no fitting & fitted        & \cite{Moore} & \\ 
\hline
\multicolumn{6}{c}{Ti$^+$} \\
 $3d^2 4p$ & $^4G^o_{5/2}$  & 28097 & 29759 & 29544 &   396~(50) \\
 $3d^2 4p$ & $^4F^o_{3/2}$  & 29401 & 30691 & 30837 &   541~(50) \\
 $3d^2 4p$ & $^4F^o_{5/2}$  & 29521 & 30813 & 30959 &   673~(50) \\
 $3d^2 4p$ & $^4D^o_{1/2}$  & 31143 & 32416 & 32532 &   677~(50) \\
 $3d^2 4p$ & $^4D^o_{3/2}$  & 31227 & 32510 & 32603 &   791~(50) \\
 $3d4s4p$  & $^4D^o_{1/2}$  & 50889 & 52185 & 52339 & -1564~(150) \\
 $3d4s4p$  & $^4F^o_{3/2}$  & 51341 &       & 52330 & -1783~(300) \\
\multicolumn{6}{c}{Ti$^{2+}$} \\
 $3d4p$    & $^3D^o_1$      & 80558 &       & 77000 & -1644~(150) \\
\end{tabular}
\end{ruledtabular}
\end{table}

\begin{figure}[htb]
\begin{center}
\includegraphics{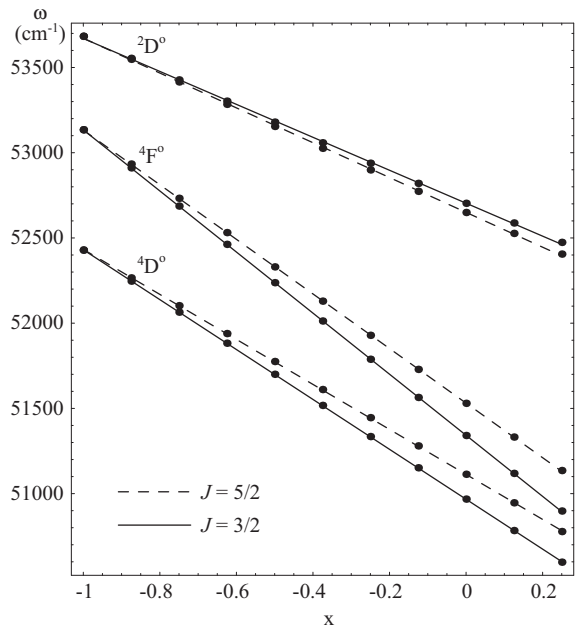}
\caption{Dependence of some odd $3d4s4p$ levels of Ti$^+$ on $x = (\alpha / \alpha_0 )^2 - 1$. Solid lines correspond to $J = 3/2$ and dashed lines to $J = 5/2$. Data points are shown from simulations, along with a linear fit that shows that there is no interaction between the transitions.
The assignment of the multiplets, from the bottom up: $^4D^o$, $^4F^o$, and $^2D^o$; the $^4D^o$, $^4F^o$ multiplets appear in the wrong order compared to experiment. The $^4D^o_{1/2}$ line is not shown since there are no nearby lines with the same angular momentum, and hence no risk of interaction. The $^4D^o_{7/2}$ and $^4F^o_{7/2, 9/2}$ lines are also omitted.}
\label{fig}
\end{center}
\end{figure}

Figure \ref{fig} shows the dependence of the highly excited $3d4s4p$ levels of Ti$^+$ on $\alpha^2$. The data points are generated from a highly saturated CI, without any fitting parameters. The dependence is very close to linear, and hence there is no level pseudo-crossing in the code. This does not completely exclude the possibility that the $^4F^o_{3/2}$ and $^4D^o_{3/2}$ levels are strongly mixed (this could be checked and accounted for if the experimental g-factors were known \cite{Dzuba2}, but they are not available). In the case of strong mixing, $q$ lies somewhere between that of the two states, thus we have increased the $^4F^o_{3/2}$ error to include this possibility.

\paragraph{Sodium ($Z=11$):} In contrast to the ions considered above,
sodium has only one external electron above closed
shells. Its ground state is $1s^2 2s^2 2p^6 3s \ ^2S_{1/2}$.
Very accurate calculations are possible for such systems by
including certain types of correlation diagrams to all orders (see, e.g.
\cite{Dzuba89,Johnson91}). However, since both relativistic and correlation
effects for Na are small we use a simplified approach. We calculate
the correlation potential $\hat \Sigma$ (the average value of this operator
is the correlation correction to the energy of the external electron) 
in the second order of MBPT, 
and use it to modify the RHF equations for the valence electron and
to calculate the so-called Brueckner-orbitals. Note that due to iterations
of $\hat \Sigma$ certain types of correlation diagrams are still included
in all orders in this procedure. The final accuracy of the energy is better
than 1\%, and for the fine structure accuracy is 2-6\% (see Table \ref{Na}).
We believe that the accuracy for the relativistic shifts $q$ is on the 
same level.

\begin{table}[htb]
\caption{Energies and relativistic energy shifts ($q$) for Na.}
\label{Na}
\begin{ruledtabular}
\begin{tabular}{llccc}
\multicolumn{2}{c}{State} & \multicolumn{2}{c}{$\omega_0$ (\cm)}
    & \multicolumn{1}{c}{$q$ (\cm)} \\
&&\multicolumn{1}{c}{theory} & experiment \cite{Moore} & \\
\hline
 $3p$ & $^2P^o_{1/2}$ & 16858 & 16956 & 45~(4) \\
 $3p$ & $^2P^o_{3/2}$ & 16876 & 16973 & 63~(4) \\
 $4p$ & $^2P^o_{1/2}$ & 30124 & 30267 & 53~(4) \\
 $4p$ & $^2P^o_{3/2}$ & 30130 & 30273 & 59~(4) \\
\end{tabular}
\end{ruledtabular}
\end{table}

\begin{table}[htb]
\caption{Energies and relativistic energy shifts ($q$) for the carbon atom and its ions (cm$^{-1}$)}
\label{carbon}
\begin{ruledtabular}
\begin{tabular}{llrrc}
\multicolumn{2}{c}{State} & \multicolumn{2}{c}{$\omega_0$ (\cm)} & $q$ (\cm) \\
&&\multicolumn{1}{c}{theory} & \multicolumn{1}{c}{experiment \cite{Moore}} & \\
\hline
\multicolumn{5}{c}{C} \\
$2s 2p^3$ & $^3D^o_3$   & 66722  & 64087 & 151~(60) \\
$2s 2p^3$ & $^3D^o_1$   & 66712  & 64090 & 141~(60) \\
$2s 2p^3$ & $^3D^o_2$   & 66716  & 64091 & 145~(60) \\     
$2s 2p^3$ & $^3P^o_1$   & 75978  & 75254 & 111~(60) \\
$2s 2p^3$ & $^3S^o_1$   &100170  &105799 & 130~(60) \\
\multicolumn{5}{c}{C$^+$} \\
$2s^2 2p$ & $^2P^o_{3/2}$ &  74  &    63 &  63~(1)  \\
$2s 2p^2$ & $^2D_{5/2}$ & 76506  & 74930 & 179~(20) \\
$2s 2p^2$ & $^2D_{3/2}$ & 76503  & 74933 & 176~(20) \\
$2s 2p^2$ & $^2S_{1/2}$ & 97993  & 96494 & 161~(30) \\
\multicolumn{5}{c}{C$^{2+}$} \\
$2s 2p$   & $^1P^o_1$  & 103955 & 102352 & 162~(20) \\
\multicolumn{5}{c}{C$^{3+}$} \\
$2p$ & $^2P^o_{1/2}$    & 64557  &  64484& 104~(20) \\
$2p$ & $^2P^o_{3/2}$    & 64686  &  64592& 232~(20) \\
\end{tabular}
\end{ruledtabular}
\end{table}

\paragraph{Carbon ($Z=6$):} Relativistic effects for carbon and its ions 
are small and calculations can be done without fitting parameters.
The ground state of neutral carbon is $1s^2 2s^2 2p^2 \ ^3P_0$.
Our RHF calculations for this atom include all electrons,
however, since we need to consider configurations with excitations 
from both $2s$ and $2p$ states, we treat both as valence states in CI.

For neutral carbon we have performed the calculations for the ground
state configuration as well as for excited configurations
$2s^2 2p3s$, $2s 2p^3$, $2s^2 2p4s$,$2s^2 2p3d$, $2s^2 2p4d$,
$2s^2 2p5d$ and $2s^2 2p6d$. However, we present in Table~\ref{carbon}
only results for the $2s 2p^3$ configuration. The relativistic energy
shift for all other configurations is small ($|q| < 100$~\cm).
This is smaller than the uncertainty in the $q$-coefficients for heavier ions. Since the analysis of quasar spectra is based on comparison of the 
relativistic effects in light and heavy atoms (ions), small relativistic energy shifts
in light atoms can be neglected. The $q$-coefficients for the $2s 2p^3$
configuration are larger because this configuration corresponds to the
$2s - 2p$ transition from the ground state. These are the lowest valence 
single-electron states with the largest relativistic effects. Other
excited configurations correspond to the $2p - ns$ or $2p - nd$ ($n \geq 3$)
transitions. However, relativistic energy shifts for higher states are smaller
\cite{Dzuba1}.

The calculations for C$^{2+}$ and C$^{3+}$ are done in the potential of the 
closed-shell (helium) core. As can be seen from Table \ref{carbon}, accuracy 
for the energies is within 10\%. We estimate the accuracy of $q$-coefficients 
at around 10\%.

\begin{table}[htb]
\caption{Energies and relativistic energy shifts ($q$) for oxygen ions.}
\label{O}
\begin{ruledtabular}
\begin{tabular}{llrrc}
\multicolumn{2}{c}{State} & \multicolumn{2}{c}{$\omega_0$ (\cm)} & $q$ (\cm) \\
&&\multicolumn{1}{c}{theory} & \multicolumn{1}{c}{experiment \cite{Moore}} & \\
\hline
\multicolumn{5}{c}{O$^+$} \\
$2s 2p^4$ & $^4P_{5/2}$   & 122620 & 119873 & 346~(50) \\
$2s 2p^4$ & $^4P_{3/2}$   & 122763 & 120000 & 489~(50) \\
$2s 2p^4$ & $^4P_{1/2}$   & 122848 & 120083 & 574~(50) \\
\multicolumn{5}{c}{O$^{2+}$} \\
$2s 2p^3$ & $^3D^o_{1}$   & 121299 & 120058 & 723~(50) \\
$2s 2p^3$ & $^3P^o_{1}$   & 143483 & 142382 & 726~(50) \\
\multicolumn{5}{c}{O$^{3+}$} \\
$2s 2p^2$ & $^2D_{3/2}$   & 129206 & 126950 & 840~(50) \\
\multicolumn{5}{c}{O$^{5+}$} \\
$1s^2 2p$ & $^2P^o_{1/2}$ &  96501 &  96375 & 309~(50) \\
$1s^2 2p$ & $^2P^o_{3/2}$ &  97091 &  96908 & 913~(50) \\
\end{tabular}
\end{ruledtabular}
\end{table}

\paragraph{Oxygen ($Z=8$):}

Relativistic effects for oxygen ions are comparatively large, 
and become larger with increasing electric charge. 
This is in agreement with semi-empirical formulae presented in \cite{Dzuba1}.
For transitions in neutral oxygen, however, $|q| < 20$~\cm; these results are not presented here.

\vspace{8pt}
This work was supported by the Australian Research Council.

\bibliography{qq}

\end{document}